\documentstyle[11pt,newpasp,twoside,epsf]{article}
\def\edcomment#1{\iffalse\marginpar{\raggedright\sl#1\/}\else\relax\fi}
\reversemarginpar

\begin{document}
\title{Properties of Proto--Planetary Nebulae}
 \author{Margaret Meixner}
\affil{University of Illinois, Dept. of Astronomy, MC-221, 
1002 W. Green St., Urbana, IL 61801}

\begin{abstract}
This review describes some general properites of proto-planetary
nebulae with particular emphasis on the recent work of
morpholgical studies.  The weight of observational evidence
shows that proto-planetary nebulae (PPNe) are most certainly axisymmetric
like planetary nebulae.  Recent work 
suggests two subclasses of PPNe optical morphology,
DUst-Prominent Longitudinally-EXtended
(DUPLEX) and Star-Obvious Low-level Elongated (SOLE).
Radiative transfer models of an example DUPLEX PPN and SOLE PPN, 
presented here, support the interpretation that DUPLEX and SOLE are
two physically distinct types of PPNe.  The DUPLEX PPNe and SOLE PPNe may
well be the precursors to bipolar and elliptical PNe, respectively.

\end{abstract}

\section{What is a Proto-Planetary Nebula?}

The proto-planetary nebula (PPN;a.k.a. post-AGB or pre-PN) stage of 
evolution immediately precedes
the planetary nebula (PN) stage.  The lifetime  for this phase is 
$\; \buildrel < \over \sim \;$1000 years and marks the time from when the 
star was forced
off the asymptotic giant branch (AGB) by intensive mass loss to when
the central star becomes hot enough (T$_{\rm eff} \sim 3\times 10^4$ K)
to photoionize the neutral circumstellar shell (Kwok 1993).  
We refer the reader to Kwok (1993) for a comprehensive review of PPN.
For short recent reviews, see Hrivnak (1997)  and van Winckel (1999).
In this short review, I summarize the basic properties of PPN
but focus primarily on the morphological studies because there
have been numerous morphological studies in the past few years and because
this particular conference is focused on morphology.

\section{General Characteristics of PPN}

PPN are like PN in that the central star illuminates a
detached circumstellar shell, however,  we observe PPN using quite
different techniques. Because PPN do not have ionized gas,
we can not use the optically bright emission lines or radio
free-free continuum commonly used for studies of PN.
Instead tracers of dust and neutral gas are employed.
In fact,  PPN are identified as stars of spectral type
B-K, luminosity class I with infrared  excesses.  These infrared excesses
arise from the circumstellar dust which was originally created
in the AGB wind.  They  
emit broad ($\sim$20 km s$^{-1}$), parabolic or double--horned 
lines  of CO rotational lines 
and OH maser lines indicative of a remnant AGB circumstellar shell.
These broad lines distinguish these PPN from pre-main sequence
and Vega-excess stars which also have infrared excesses but typically
narrower or non-existent molecular lines.

Candidates for PPN are discovered using infrared sky surveys.
One of the most famous PPN, AFGL 2688 (a.k.a the Egg  Nebula),
was discovered by Ney et al. (1975) in an infrared sky survey
done by the airforce. Studies using the IRAS all sky survey
have used two approaches to identify candidates. 
One way is to use IRAS colors ([25]-[60] vs. [12]-[25]),
mark the locations of known AGB stars and PN and choose PPN candidates
from the regions in between  (van der Veen, Habing \& Geballe 1989;
Hrivnak, Kwok \& Volk 1989; Hu et al. 1993).
The second way is to cross correlate the IRAS catalog with optical
star catalogs and choose objects in common (Oudmaijer et al. 1992).

Several initial followup studies focused on ground based photometry 
observations and models of the spectral
energy distributions (SEDs) of these PPN candidates.  Van der Veen et al.
(1989) identified four types of SEDs which they attributed to
optical depth differences in the dust shells. Type I has a flat spectrum
from 4 to 25 $\mu$m with a steep fall off at short wavelengths.
Type II have a maximum near 25 $\mu$m and a gradual fall-off to shorter
wavelengths. Type III have a maximum near 25 $\mu$m, a steep fall off
at shorter wavelengths to a plateau between 1 and 4 $\mu$m.
Type IV have two distinct maxima, one near 25 $\mu$m and the second
at $\lambda < 2$ $\mu$m.   Hrivnak \& Kwok (1991) suggested
that the 21$\mu$m PPN were quite similar to objects like AFGL 2688
except for viewing angle based on the differences in their SEDs.
The 21$\mu$m PPN have a Type IV SED while AFGL 2688  has a Type III  SED
and we could be viewing the 21$\mu$m PPN down the poles while we
view AFGL 2688  edge-on.

\section{ Morphologies}

The study of PPN morphologies offers us insight on the axisymmetric
PN issue because they are the missing link between two well studied
groups: PN and AGB stars.  Moreover, their morphologies are relatively
pristine  fossil records of the AGB mass loss process because 
shaping by the hot, fast stellar wind of a PN nucleus has 
probably not occured (see Schonberner in this volume). Hence we
can determine ``initial conditions'' for the interacting winds models. 
Figure 1 shows our working model for a PPN circumtellar shell based
on the observed evidence that most PN are axially symmetric (e.g. Balick 1987)
while the outer shells of AGB circumstellar shells are spherically
symmetric (e.g. Neri et al. 1998).  In the PPN fossil record, radial
distance directly corresponds to time. The maximum
radius, $\rm R_{max}$ marks when the mass loss began.  As we move
inwards, we see the spherically symmetric shell created by the
AGB wind. The superwind radius, $\rm R_S$, corresponds to when
the mass loss began
to increase in rate and to assume an  axial as opposed to spherical
symmetry. We note that our use of the term superwind is slightly modified
from the intention of Iben \& Renzini (1983) in that we include the
symmetry change in addition to an increase in mass loss rate.
The inner radius, $\rm R_{in}$
marks when the mass loss stopped and the size of the inner radius reflects
the dynamical age that has passed since the star left the AGB (Meixner
et al. 1997).

Studies of PPN morphologies have used tracers of dust  (thermal
infrared radition or optical/near-infared scattered starlight) or
molecular gas (CO maps or maser maps).  Since the molecular gas
observations are covered by others in this volume (see Huggins; Alcolea;
Fong et al.), I will focus on the tracers of dust.  In PPN,
the central star heats the dust and the dust both scatters the starlight
and radiates in the thermal infrared ($\lambda > 5 \mu$m).  
In our working model, we expect that the scattered starlight will
preferentially leak out of the lower density bipolar openings of the
dustshell and will be the most intense in the inner regions where both
the starlight and dust density are highest.  In the thermal infrared,
what we see depends on the wavelength we observe because dust radiates
as a modified black body and hence depends sensitively on temperature.
In these dust shells, the temperature decreases from 
at $\sim$200 K at the inner radius  to $\sim$30 K at the outer radius.
Mid-infrared (Mid-IR; 8-25$\mu$m) 
emission arises exclusively from the inner regions
where the dust is warm.  Far-IR and submillimeter
radiation ($>$50 $\mu$m) arises from the outer regions as well as the
inner regions, however, the angular resolution at these wavelengths is 
presently quite poor (10-40\arcsec) which prevents investigation of the inner,
axisymmetric regions.  

\begin{figure}
\plotfiddle{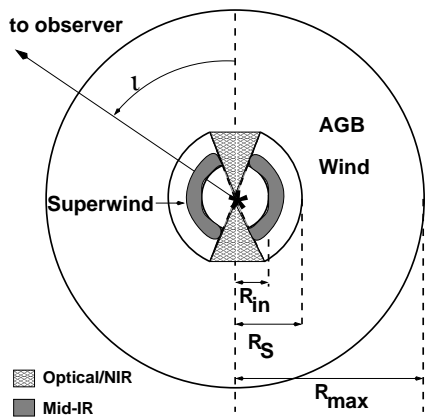}{2.0in}{0}{35}{35}{-70}{0}
\caption{Schematic representation of our working model of a 
Proto-Planetary Nebula Dustshell.}
\end{figure}

\subsection{ Mid-IR Imaging Studies}

Ground based mid-IR imaging studies of PPN have had typical angular
resolutions of about 1\arcsec~ and enough spectral resolution to
separate dust features and dust continuum.  The larger telescopes
coming on line, e.g. Keck, VLT, Gemini and MMT, promise diffraction limited
performance as good as 0.\arcsec 2  (e.g. see Morris in this volume,
and Jura \& Werner 2000).
A number of published  mid-IR imaging studies of PPN have
focused on one to five well resolved PPN and 
usually include radiative transfer
modeling (Skinner et al. 1994; Hawkins et al. 1995; Hora et al. 1996;
Dayal et al. 1998; Meixner et al. 1997; Skinner et al. 1997).   
A recent survey paper of 66 PPNe (Meixner et al. 1999) 
provides the largest data base of PPNe candidate mid-IR images to date.  
It also  incorprates  the results of previously published works.  
Considering all the published mid-IR images to date, there are three
main points that can be made (Meixner et al. 1999).  
First, of the 73 PPNe candidates, 33\% have
been resolved with $\sim$1\arcsec~ resolution.  The cooler and
brighter objects are easier to resolve probably because cooler shells
are more distant from the central star and hence larger and brighter
shells are either closer or more luminous which create larger mid-IR 
emission regions.  Second, all of the well resolved PPNe are axisymmetric.
Thirdly,  there appear to be two morphological types in the well-resolved
mid-IR PPNe candidates which we have called toroidal  and core/elliptical.
Toroidals, as exemplified by IRAS 07134+1005 (Fig. 2), 
have elliptical/round outer
perimeters and two peaks which can be interpreted as limb brightened
peaks of an equatorial density enhancement.  The central star usually appears
between the two peaks.
Core/ellipticals, as exemplified by the Red Rectangle (Fig. 2), 
have tremendously
bright, unresolved cores at their centers and low surface brightness
elliptical shaped nebulae surrounding these cores. The extension of
the low surface brightness emission is in the same direction as the
optical reflection nebulosity found in these objects.  

\begin{figure}
\plotfiddle{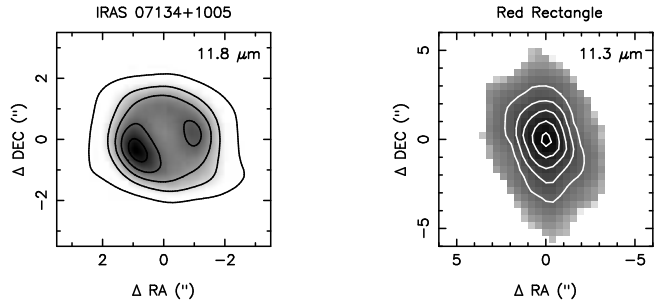}{1.5in}{0}{80}{80}{-150}{-20}
\caption{Mid-IR images of two types of PPNe with the wavelengths in the
upper right corner. Left: IRAS 07134+1005, an
example of a Toroidal PPN (Meixner et al. 1997). Right:  Red Rectangle,
an example of a core/elliptical (Meixner et al. 1999).}
\end{figure}

\subsection{Optical Polarimetry and Imaging Studies}

The optical and near-IR polarimetry and imaging observations of 
the scattered starlight in PPN predate (1970's)
and far out number these mid-IR studies. Here I only have room to summarize
some recent work. Two large survey polarimetry studies 
of PPN have revealed a large amount of polarization at the PPN stage 
which indicates that the  PPN stage is axisymmetric.  Using broad band
polarimetry, Johnson \& Jones (1991)
investigated 38 objects ranging from the AGB to PPN to PN stages 
and found that the polarization increased from the AGB to PPN stage
and then decreased in the PN stage. Using spectropolarimetry, Trammell,
Dinerstein \& Goodrich (1994), studied 31 PPN and found 80\% had
some intrinsic polarization.  They also classified polarized PPN into
Type 1 and Type 2.  Both have large polarizations, but Type 1's also have
a large position angle rotation with wavelength which suggests that
Type 1's may be more bipolar in shape.  See Gledhill et al. and
Su et al. in this volume for recent near-IR polarimetry work on PPN.

High angular resolution 
optical and  imaging studies of PPN are numerous and have exploded
with use of HST because the compact nature of PPNe is well suited for
HST.   
A number of the studies focus on one or a few objects and range
from phenomenological discussions to quantitative modelling in their
interpretations (e.g. Sahai, Bujarrabal \& Zijlstra 1999; Sahai et al. 1999; 
Kwok, Su \& Hrivnak 1998; Su et al. 1998; Sahai et al. 1998; Kwok et al. 1996;
Skinner et al. 1997; Trammell \& Goodrich 1996;
Bobrowsky et al. 1995; Latter et al. 1993; 
in this volume see  Trammell et al., Hrivnak et al., Bobrowsky et al.
and Bieging et al.). Two recent papers cover
a significant number of PPN and strive for a more global picture of PPN.
Hrivnak et al. (1999) pursued a ground based imaging study of 10 PPN
with angular resolutions of $\sim$0.\arcsec 75.
Ueta, Meixner \& Bobrowsky (2000) imaged 27 PPN using the HST WFPC2 with
angular resolutions of 0.\arcsec 046 (see also Ueta et al. in this volume).
If we combine the imaging results from all these papers, we find that
80\% of the 44 PPN have resolvable optical reflection nebulosities
and all of the well resolved reflection nebulosities are axisymmetric.
While many of these studies have focused on PPN with obscured central
stars, the Ueta et al. (2000) survey included many PPN with optically
visible, prominent stars and they discovered two types of optical
morphology in the PPN.

\section{Two Types of PPN}

\begin{figure}
\plotfiddle{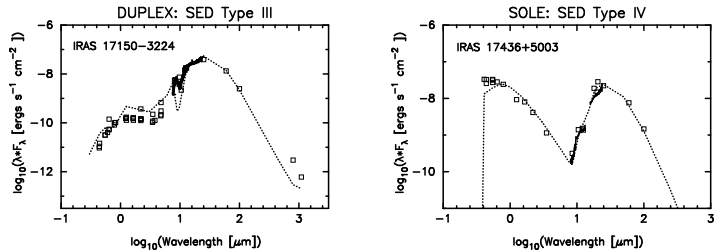}{1.0in}{0}{80}{80}{-150}{-30}
\caption{The spectral energy distributions for DUPLEX PPN,
IRAS 17150-3224 (left) and for SOLE PPN IRAS 17436+5003 (right).
Photometry data are the squares, spectroscopy the solid lines
and  the model in dashed lines. From Meixner et al. (2000).}
\end{figure}

The two types of PPN have been called DUst-Prominent Longitudinally-EXtended
(DUPLEX) and Star-Obvious Low-level Elongated (SOLE) PPN  (see Ueta et al.
in this volume).
These names describe their optical morphological appearance and 
their acronyms describe the two lobed structures seen 
in DUPLEX PPNe and the continuous structures seen in SOLE PPNe.
Besides their optical appearances, DUPLEX and SOLE PPNe differ in
their mid-IR morphologies: DUPLEX are core-ellipticals, while SOLE
are toroidals.  They also have distinctly different
SEDs:  DUPLEX have type II or III SEDs, while
SOLE have type IV SEDs in the van der Veen et al. 1989 classification.
Ueta et al. (2000) claim that the cause of these differences is the
optical depth of the dust shell.  SOLE nebulae have less dust optical
depth than DUPLEX nebulae and, hence, the central star is visible no
matter the inclination angle.  They further suggest that DUPLEX PPNe
may well be the precursors of bipolar PNe while SOLE PPNe may be
the precursors to the elliptical PNe based on their differences
in morphologies and galactic height distributions.

This interpretation is controversial, judging by the avid discussion after 
my talk.  The other point of view, presented by Hrivnak (in this volume)
is that these optical morphological differences are due primarily to
inclination angle differences.  That is, all these PPNe are the same physical
type of beast, just viewed from different angles on the sky.
The best way to resolve such controversy is to make radiative transfer
models using axially symmetric dust codes to derive optical depths,
and structures for all of the PPNe and compare their derived properties.
Here, we present model results of one DUPLEX PPN, IRAS 17150-3224, and one
SOLE PPN, IRAS 17436+5003, to demonstrate that these two sources, which
are among the best examples of their classes, are physically quite different.
We use a radiative transfer code that we used in Meixner et al. (1997)
and  Skinner et al. (1997).  Su et al. (1998 and in this volume) are using 
a different code with similar aims but have concentrated on primarily on 
DUPLEX PPN so far.   

\begin{figure}
\plotfiddle{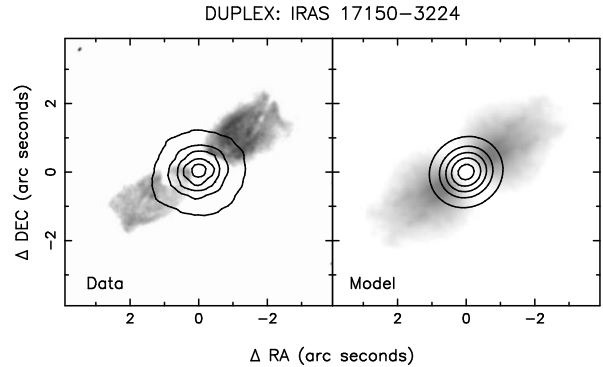}{1.7in}{0}{80}{80}{-160}{-30}
\caption{The model images and data for DUPLEX PPN,
IRAS 17150-3224.
Grayscale  is the HST B band images  and the contours are
the 9.8$\mu$m mid-IR images. From Meixner et al. (2000). }
\end{figure}

\begin{figure}
\plotfiddle{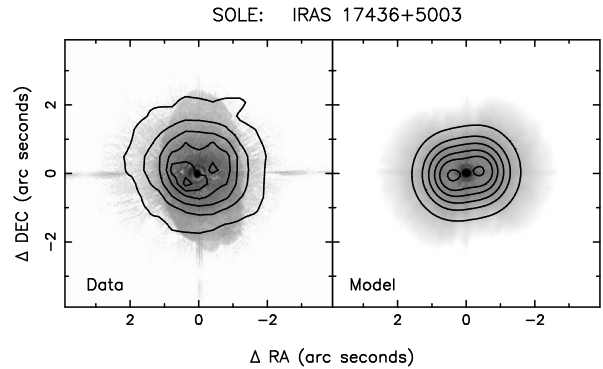}{1.7in}{0}{80}{80}{-160}{-30}
\caption{The model images and data for for SOLE PPN IRAS 17436+5003.
Grayscale  is the HST V band images  and the contours are
the 12.5$\mu$m mid-IR images. From Meixner et al. (2000). }
\end{figure}

We have constrained the model using our HST and mid-IR images
and photometry from the literature.  A full discussion of these models
will appear in Meixner, Ueta \& Bobrowsky (2000).  Comparison of
the model images and SED data demonstrates
a reasonable match of the model with the data (Figs. 4 and 5).  The derived
parameters from the models appear in the Table and reveal the 
physical reasons for the apparent morphological differences.
First, both objects are best fit by a 90$^\circ$ inclination
angle; i.e. we are viewing both edge-on. Thus, we are not viewing
the SOLE PPN, IRAS 17436+5003, near the pole which  would be
expected if our viewing angle were the main cause of the morphological
differences.  We note that other PPN in the Ueta et al. (2000) sample
show qualitative evidence for inclination angles different than 90$^\circ$;
e.g. unbalanced lobes for DUPLEX and less elliptical nebulae for SOLE.
Second, the optical depth for the DUPLEX PPN, IRAS 17150-3224, is
significantly higher than for the SOLE PPN, IRAS 17436+5003
and explains why we do not see the central star in the former but do
in the latter.  The cause for the difference in optical depth is
the history of mass loss.  IRAS 17150-3224 experienced a more
intensive mass loss rate than IRAS 17436+5003 that resulted in
a denser dust cocoon of significantly higher mass. Both this
higher mass and the higher luminosity for this source suggest
that IRAS 17150-3224 originated from a higher mass star.  
These results are, of course, distance dependent and the distance
maybe uncertain by a factor of two.  However, the optical
depth is distance independent and even with the distance uncertainty
it seems reasonable to conclude that IRAS 17150-3224 is more luminous
and had a higher mass progenitor.

\begin{table}
\begin{tabular}{lccccccc}
\hline
Object &  $\tau_{eq,9.7\mu m}$&  incl. & \.M$_{sw}$ ($\rm M_\odot yr^{-1}$) & 
L$_*$ (L$_\odot$) & D (kpc)  \\ \hline
DUPLEX: \\
IRAS 17150-3224 &  1.8  & 90$^\circ \pm$5 & $50\times 10^{-5}$ &  27000 & 3.6  \\
 SOLE: \\
IRAS 17436+5003 & 0.9 &  90$^\circ \pm$20&  $6\times 10^{-5}$  & 3900 & 1.2  \\
\end{tabular}
\end{table}

\section{Summary points}

The observational evidence from a number of independent studies
clearly shows that PPNe  are intrinsically  axisymmetric.
Thus the axisymmetry that we observe in PNe must predate the PPNe
stage.  Most likely the axisymmetry originates at the end of the
AGB phase, because observations of the outer regions of AGB envelopes
show a spherical symmetry.  With the variety of PNe morphologies discussed
at this conference (e.g.  round, elliptical, bipolar and point symmetric),
we must now  begin to ask: Do we see examples of PPNe with
these corresponding subtle differences in morphologies?
I think we are beginning to see these differences.  The DUPLEX and
SOLE PPNe may well be the precursors for bipolar and elliptical PNe.

\end{document}